\documentclass[useAMS, usenatbib, amssymb]{mn2e}
\usepackage{psfig}
\usepackage{graphicx}
\usepackage{epsf}
\usepackage{bm}
\usepackage{lscape}

\title [Transformations between 2MASS, SDSS and BVRI photometric systems]
{Transformations between 2MASS, SDSS and BVRI photometric systems:
bridging the near infrared and optical}

\author[Bilir et al.]
       {S. Bilir,${^1} \thanks{E-mail: sbilir@istanbul.edu.tr}$
        S. Ak$^{1}$, S. Karaali$^{2}$, A. Cabrera-Lavers$^{3,4}$, T. S. Chonis$
^{5}$, C. M. Gaskell$^{5,6}$
\\
  $^1$Istanbul University Science Faculty, Department of Astronomy and Space 
Sciences, 34119, University-Istanbul, Turkey\\
  $^2$Beykent University, Faculty of Science and Letters, Department of Mathematics  
and Computer, Beykent 34398, Istanbul, Turkey\\
  $^3$Instituto de Astrof\'{\i}sica de Canarias, E-38205 La Laguna, Tenerife, Spain\\
  $^4$GTC Project Office, E-38205 La Laguna, Tenerife, Spain\\
  $^5$Department of Physics \& Astronomy, University of Nebraska, Lincoln, NE  
68588-0111, USA\\
  $^6$Present Address: Department of Astronomy, University of Texas, Austin, 
TX 78712-0259, USA\\}

\date{Accepted 2007 month day.
Received year month day; }

\pagerange{\pageref{firstpage}--\pageref{lastpage}} \pubyear{2007}

\begin{document}

\maketitle

\label{firstpage}

\begin{abstract}
We present colour transformations for the conversion of the {\em
2MASS} photometric system to the Johnson-Cousins $UBVRI$ system and
further into the {\em SDSS} $ugriz$ system. We have taken {\em SDSS}
$gri$ magnitudes of stars measured with the 2.5-m telescope from
$SDSS$ Data Release 5 (DR5), and $BVRI$ and $JHK_{s}$ magnitudes
from Stetson's catalogue and \citet{Cu03}, respectively. We
matched thousands of stars in the three photometric systems by their
coordinates and obtained a homogeneous sample of 825 stars by the
following constraints, which are not used in previous
transformations: 1) the data are de-reddened, 2) giants are omitted,
and 3) the sample stars selected are of the highest quality. We give
metallicity, population type, and transformations dependent on two
colours. The transformations provide absolute magnitude and distance
determinations which can be used in space density evaluations at
short distances where some or all of the {\em SDSS} $ugriz$
magnitudes are saturated. The combination of these densities with
those evaluated at larger distances using {\em SDSS} $ugriz$
photometry will supply accurate Galactic model parameters,
particularly the local space densities for each population.

\end{abstract}


\section{Introduction}

Among several large surveys, two have been most widely used in
recent years. The first, the Sloan Digital Sky Survey
\citep[$SDSS$;][]{York00}, is the largest photometric and
spectroscopic survey in optical wavelengths. Secondly, the Two
Micron All Sky Survey \citep[$2MASS$;][]{2mass06} has imaged the sky
across infrared wavelengths. 

{\em SDSS} obtains images almost simultaneously in five broad bands
($u$, $g $, $r$, $i$, and $z$) centered at 3540, 4760, 6280, 7690
and 9250 \AA, respectively \citep{Fukugita96, Gunn98, Hogg01,
Smith02}. The photometric pipeline \citep{Lupton01}
detects the objects, matches the data from the five filters, and
measures instrumental fluxes, positions and shape parameters. The
shape parameters allow the classification of objects as ``point
source'' (compatible with the point-spread function) or
``extended''. The magnitudes derived from fitting a point-spread
function (PSF) are currently accurate to about 2 per cent in $g$,
$r$, and $i$, and 3--5 per cent in $u$ and $z$ for bright ($<2 0$
mag) point sources. Data Release 5 (DR5) is almost 95 per cent
complete for point sources to ($u$, $g$, $r$, $i$, $z$)=(22, 22.2,
22.2, 21.3, 20.5). The median full-width at half-maximum of the PSFs is
about 1.5 arcsec \citep{Abazajian04}. The data are saturated at
about 14 mag in $g$, $r$, and $i$, and about 12 mag in $u$ and $z$
\citep[see, for example,][]{Chonis07}. 

{\em 2MASS} provides the most complete database of near infrared
(NIR) Galactic point sources available to date. During the
development of this survey, two highly automated 1.3-m telescopes
were used: one at Mt. Hopkins, Arizona to observe the Northern Sky,
and the other at Cerro Tololo Observatory in Chile to complete the
survey's Southern half. Observations cover 99.998 per cent
\citep{2mass06} of the sky with simultaneous detections in $J$ (1.25
$\mu $m), $H$ (1.65 $\mu$m), and $K_{s}$ (2.17 $\mu$m) bands up to
limiting magnitudes of 15.8, 15.1, and 14.3, respectively. Bright
source extractions have 1$\sigma$ photometric uncertainty of $<$
0.03 mag and astrometric accuracy on the order of 100 mas.
Calibration offsets between any two points in the sky are $< 0.02$
mag. The passband profiles for $BVRI$, $ugriz$, and $JHK_{s}$ photometric 
systems are given in Fig. 1.

\begin{figure}
\begin{center}
\includegraphics[angle=0, width=80mm, height=98.7mm]{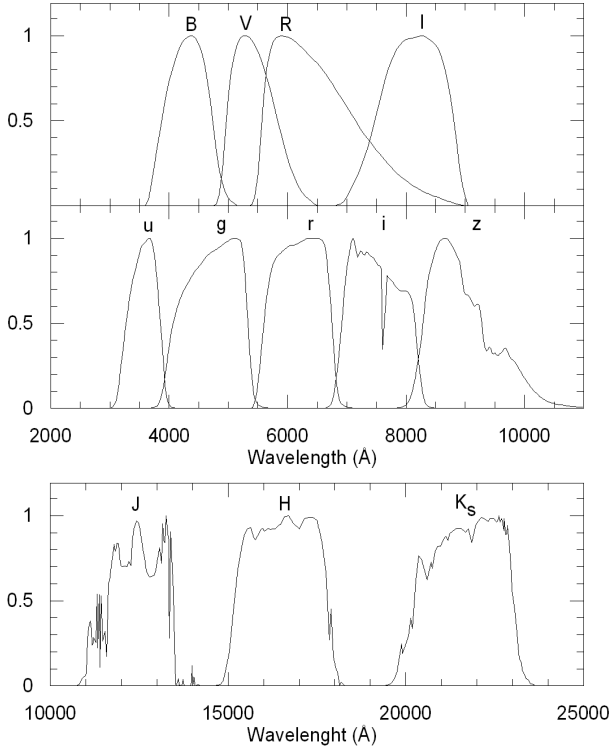}
\caption[] {Normalized passbands of the Johnson-Cousins $BVRI$
filters (upper panel), the {\em SDSS} $ugriz$ filters (middle
panel), and the {\em 2MASS} filters (lower panel).}
\end{center}
\end{figure}

It is important to derive transformations between a newly defined
photometric system and those that are more traditional (such as the
Johnson-Cousins $UBVRI$ system). A number of transformations 
between $u'g'r'i'z'$, $ugriz$ and $UBVR_{C}I_{C}$ exist. The 
$u'g'r'i'z'$ system is referred to the similar filter system used 
on the 0.5-–m Photometric Calibration Telescope at {\em SDSS}.
It should be noted that there are differences between the 
$u'g'r'i'z'$ and $ugriz$ systems. These are discussed in \cite{Tucker06}, 
\cite{Davenport07}, and \cite{Smith07}. In this paper we are 
concerned with transformation to and from the $ugriz$ system of 
the 2.5-m. The first transformations derived between the {\em SDSS} 
$u'g'r'i'z'$ system and the Johnson–-Cousins photometric system were 
based on observations in $u'$, $g'$, $r'$, $i'$, and $z'$ filters 
\citep{Smith02}. In their work, the standards refer to the {\em SDSS} 
filter-detector combination used at the 1.0-m telescope at the US Naval 
Observatory. There are slight zero-point differences between these and 
the filters used at the 2.5-m telescope at Apache Point Observatory (APO) 
which lead to systematic differences between the magnitudes evaluated at 
the two observatories \citep{Rider04}.

An improved set of transformations between the observations obtained
in $u'g'r'$ filters at the Isaac Newton Telescope (INT) at La Palma,
Spain, and the \citet{Landolt92} {\em UBV} standards is derived by
\citet*{KBT05}. The INT filters were designed to reproduce the {\em
SDSS} system. \citet{KBT05} presented for the first time
transformation equations depending on two colours. 

\citet{Rodgers06} considered two--colour or quadratic forms in their 
transformation equations. \citet*{Jordi06} used {\em SDSS} DR4 and $BVRI$
photometry taken from different sources and derived population (and
metallicity) dependent transformation equations between {\em SDSS}
and $UBVRI$ systems. They also give transformation equations
between the {\em SDSS} and {\em RGU} systems. {\em RGU} 
is a photographic system founded by \citet{Becker1938}, and used
in the Basle Halo Program \citep{Becker1965} which presented the largest
systematic survey of the Galaxy. The most recent work is
by \citet{Chonis07} who used transformations from {\em SDSS} $ugriz$
to {\em UBVRI} not depending on luminosity class or metallicity to
determine CCD zero points.

The first transformations between {\em 2MASS} and other photometric 
systems are those of \citet{Walkowicz04} and \citet{West05} who 
determined the level of magnetic activity in M and L dwarfs. The aim 
of \citet{Davenport06} in deriving equations between {\em 2MASS} and 
other systems was to estimate the absolute magnitudes of cool stars. 
\citet{Covey07} consider the $ugrizJHK$ stellar locus (i.e., the
position of main-sequence stars in the 7-dimensional colour diagram)
and show how it can be used to identify objects with unusual colours.
Our aim in the present paper is to derive transformation equations
between {\em 2MASS}, {\em SDSS} and {\em BVRI} which can be used for
various applications. Our hope is that such equations may help
Galactic researchers combine {\em 2MASS} and {\em SDSS} data in
model parameter estimations. Thus, {\em 2MASS} data would be used
to fill the space density gap at short distances where {\em SDSS}
data are saturated. We will use all the procedures described in
recent works which improve the transformations between pairs of
systems to derive the most accurate transformation equations between
the three photometric systems previously mentioned. Our
transformation equations will be based on the currently available data
\citep[DR5;][]{Adelman-McCarthy07}, and they will be dependent on
luminosity, metallicity and two colours.

In Section 2 we present the sources of our star sample and the
criteria applied to the chosen stars. The transformation equations
are given in Section 3 . Finally, in Section 4, we discuss our
results.

\section{Data}

The data used for our transformations were taken from sources
previously discussed. The first main source of our data was the
Stetson Catalogue. Stetson used a large set of multi-epoch CCD
observations centered on Landolt fields and other regions in the sky
and reduced them in a homogeneous manner tied to the Landolt $UBV
RI$ standards. The larger area coverage and greater sensitivity of
the CCD observations compared to the earlier photomultiplier
observations permitted Stetson to include stars down to $V\sim20$
mag. Since 2000, Stetson has been publishing a gradually growing
list of suitable faint stars \citep{Stetson00} with repeat
observations which can be found at the website of the Canadian
Astronomy Data
Center\footnote{http://www2.cadc-ccda.hia-iha.nrc-cnrc.gc.ca/community/STETSON/archive/}.
The Stetson catalogue contains only stars that were observed at
least five times under photometric conditions with the standard
error of the mean magnitude less than 0.02 mag in at least two of
the four filters. Stetson's database also contains fields not
covered by Landolt (e.g., fields in globular clusters and in nearby
resolved dwarf galaxies). While Landolt's original fields contained
mainly Population I stars, Stetson's new fields also include a
sizable fraction of Population II stars. Since Stetson's catalogue
does not include $U$-band photometry, we derive transformations for
$BVRI$ only. The available form of the catalogue has 40,090 stars.

The second source we use is {\em SDSS}
DR5\footnote{http://www.sdss.org/dr5/}. We selected the relevant
standard star sample by matching Stetson's published photometry to
{\em SDSS} DR5 photometry. Matching of the stars was done with
Robert Lupton's SQL code which is published on the {\em SDSS} DR5
website. We obtained 3,798 stars by matching Stetson stars and {\em
SDSS} DR5 stars.

The last source for our work is the {\em 2MASS All-Sky Catalog of
Point Sources} \citep{Cu03}. {\em 2MASS} is not as deep of a survey
as {\em SDSS}. Thus, only 1,984 out of the 3,798 stars overlapped
with the Stetson data and the {\em SDSS} DR5 data. {\em 2MASS}
magnitudes are adopted from
SIMBAD\footnote{http://simbad.u-strasbg.fr/simbad/}.

The near infrared magnitudes of the 1,984 stars found in all three
photometric systems are not as sensitive as their optical
magnitudes. To select the more sensitive near infrared magnitudes of
the selected stars, we used the magnitude flags, labelled ``AAA'',
which indicates the quality of the magnitudes for the three filters in
the {\em 2MASS} All-Sky Catalog of Point Sources \citep{Cu03}.
After applying this selection criterion based on the quality of 
the data, the total number of sufficient stars in all three photometric 
systems was reduced to 886.

\subsection{Reddening}

The standard stars lie in fields with different Galactic latitudes,
thus, each field has a different amount of reddening. Some stars 
in the sample are within 150 pc and therefore should not affected by 
reddening from Galactic dust. The $E(B-V)$ colour excesses
of stars have been evaluated in two steps. First, we used the maps
of \citet*{Schlegel98} and evaluated an $E_{\infty}(B-V)$ excess for
each star. We then reduced them by the following procedure
\citep{Bahcall80}:

\begin{equation}
A_{d}(b)=A_{\infty}(b)\Biggl[1-exp\Biggl(\frac{-\mid
d~sin(b)\mid}{H}\Biggr)\Biggr].
\end{equation}
Here, $b$ and $d$ are the Galactic latitude and distance of the
star, respectively. $H$ is the scaleheight for the interstellar
dust which is adopted as 125 pc \citep{Marshall06} and
$A_{\infty}(b)$ and $A_{d}(b)$ are the total absorptions for the
model and for the distance to the star, respectively. $A
_{\infty}(b)$ can be evaluated by means of Eq. (2):

\begin{equation}
A_{\infty}(b)=3.1E_{\infty}(B-V).
\end{equation}
$E_{\infty}(B-V)$ is the colour excess for the model taken from the
NASA Extragalactic
Database\footnote{http://nedwww.ipac.caltech.edu/forms/calculator.html}.
Then, $E_{d}(B-V)$, i.e. the colour excess for the corresponding
star at the distance $d$, can be evaluated by Eq. (3) adopted for
distance $d$,

\begin{equation}
E_{d}(B-V)=A_{d}(b)~/~3.1.
\end{equation}

We have omitted the suffixes ${\infty}$ and $d$ from the colour
excess $E(B-V)$ in our tables and figures. However, we use the 
terms ``model'' for the colour excess of \citet{Schlegel98} and ``reduced'' 
for the colour excess corresponding to distance $d$. The total absorption
$A_{d}$ used in this section and the classical total absorption
$A_{V}$ have the same meaning. As shown in Fig. 2, there are no
differences between the model $E(B-V)$ colour excesses and the
reduced ones for heights larger than $z\sim0.5$ kpc above the
Galactic plane. Here, $z=d\sin(b)$, where the distance $d$ is
evaluated by the combination of the apparent $g$ and absolute $M_{g}$
magnitudes of a star, i.e. $g-M_{g}=5\log d-5+A_{g}$, where $A_{g}$ is 
the total absorption. The absolute magnitude of stars with 
$4<M_{g}\leq8$ were determined by the procedure of \citet{KBT05}, 
whereas for $M_{g}>8$, we followed the procedure 
of \citet*{Bilir05}.

\begin{table*}
{\scriptsize
\center
\caption{Johnson-Cousins, {\em SDSS}, and {\em 2MASS} magnitudes of
the sample stars (825 total stars). The columns give: (1) Star name;
(2) and (3) Galactic coordinates; (4) $V$--apparent magnitude; (5)
and (6) $(B-V)$ and $(R-I)$ colour indices; (7) $g$--apparent magnitude;
(8), (9), and (10) $(u-g)$, $(g-r)$, $(r-i)$ colour indices; (11)
$J$-apparent magnitude; (12) and (13) $(J-H)$, $(H-K_{s})$ colour
indices, and (14) reduced $E_{d}(B-V)$ colour excess.  The complete
table is available in electronic format.}
\begin{tabular}{lccccccccccccc}
\hline
(1) & (2) & (3) & (4) &  (5) &  (6) & (7) &  (8) &  (9) &  (10) &    (11) &
 (12) &  (13) & (14) \\
Star & $l~(^{\circ})$ & $b~(^{\circ})$ & $V$ &  $(B-V)$ &  $(R-I)$ & $g$ &  
$(u-g)$ &  $(g-r)$ &  $(r-i)$ &  $J$ &  $(J-H)$ &  $(H-K_{s})$ & $E_{d}(B-V)$ \\
\hline
L107-S61& 5.437 & 41.308 & 18.472 & 1.499 & 1.054 & 19.334 & 2.695 & 1.466 & 0.796 & 15.357 & 0.775 & -0.100 & 0.093 \\
L107-S83& 5.454 & 41.266 & 17.421 & 1.548 & 1.259 & 18.272 & 2.736 & 1.468 & 1.015 & 13.862 & 0.672 &  0.262 & 0.076 \\
L107-S97& 5.499 & 41.257 & 16.145 & 0.692 & 0.461 & 16.446 & 1.267 & 0.507 & 0.176 & 14.783 & 0.397 & -0.081 & 0.103 \\
 . &          . &          . &          . &          . &          . &          . &      
    . &          . &          . &          . &. &          . &          . \\
         . &          . &          . &          . &          . &          .
&          . &          . &          . &          . &          . &
. &          . &          . \\
         . &          . &          . &          . &          . &          .
&          . &          . &          . &          . &          . &
. &          . &          . \\
\hline
\end{tabular}}
\end{table*}

\begin{figure}
\begin{center}
\includegraphics[angle=0, width=80mm, height=51.6mm]{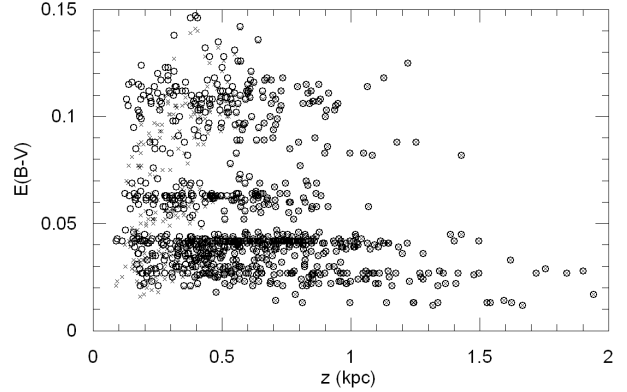}
\caption[] {$E(B-V)$ colour-excess versus height from the Galactic
plane, $z$. The symbol ($\circ$) indicates the colour excess
estimated by the procedure of \citet{Schlegel98}. ($\times$)
corresponds to the reduced colour excess \citep{Bahcall80}.}
\end{center}
\end{figure}

In order to determine total absorptions, $A_{m}$, for the {\em SDSS} bands,
we used $A_{m}/A_{v}$ data given by \citet{Fan1999}, i.e. 1.593, 1.199, 
0.858, and 0.639 for $m$=$u$, $g$, $r$, and $i$, respectively.

We used the equation of \citet{Cardelli89} for de-reddening the
$R-I$ colour and those of \citet{Fiorucci03} \citep[see also;][]
{BilirGuverAslan2006, Aketal2007} for the {\em 2MASS}
magnitudes:

\begin{equation}
E(R-I)= 0.60 E(B-V),
\end{equation}

\begin{equation}
A_{J} = 0.887 E(B-V),
\end{equation}

\begin{equation}
A_{H} = 0.565 E(B-V),
\end{equation}

\begin{equation}
A_{K_{s}}= 0.382 E(B-V).
\end{equation}
All the colours and magnitudes with subscript ``0'' will be mentioned 
as de-reddened ones, hereafter.

The Galactic coordinates of the Stetson fields and the corresponding
$E(B-V)$ colour-excesses are given in Fig. 3. The field with
$E(B-V)>0.4$ is omitted.

\begin{figure}
\begin{center}
\includegraphics[angle=0, width=80mm, height=60.9mm]{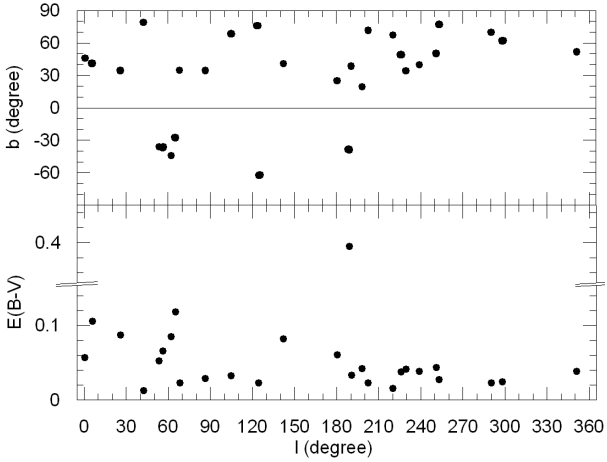}
\caption[] {Galactic coordinates of the Stetson fields and the
corresponding $E(B-V)$ colour-excess of \citet{Schlegel98}.}
\end{center}
\end{figure}

\subsection{Separation of dwarfs and giants}
The $(u-g)_{0}-(g-r)_{0}$ two-colour diagram (Fig. 4) of stars
common in all three catalogues and with the best quality $J$, $H$,
and $K_{s}$ magnitudes indicate that additional constraints are
necessary to obtain a homogeneous star sample. There is an
unexpected scattering and the stellar locus is wide. We adopted red
stars (16 in total), with $(u-g)_{0}>2.85$ mag which lie below the
concentrated stellar locus as giants. Additionally, we identified 11
metal rich giants by the procedure of \citet{Ivezic07} and omitted
them from the sample. These authors calibrated the $(u-g)_{0}-
(g-r)_ {0}$ two-colour diagram with metallicity ($[M/H]$) and surface 
gravity ($\log g$) which provides dwarf–-giant separation. The 11 stars 
mentioned above lie at the left of the locus with $\log g=2$ and 
$[M/H]=0$ dex, based on the \citet{Kurucz79} stellar model. Finally, 
we used the following procedure of \citet{Helmi03}, and identified 33 
stars as metal – poor giants: This procedure is based on the Spaghetti Photometric 
Survey (SPS) \citep{Morrison00} with Washington photometry where 
metal poor stars could be isolated on the basis of (M-T2) and (M-51) 
colours, sensitive to temperature and strength of the Mgb and MgH 
features near 5200$\AA$, respectively; and a first estimate of their 
luminosity classes were obtained. The candidates were observed 
spectroscopically and were classified into dwarfs and giants using 
the following indicators \citep{Morrison2003}: 1) The Mgb and MgH 
features near 5200$\AA$, which are characteristic of dwarfs and are 
almost absent in giant stars for $0.8\leq(B-V)\leq1.3$. 2) The 
CaI 4227$\AA$ line, which is usually present in dwarfs and absent in 
giants; may be visible in metal--poor giants with $[M/H]<-1.5$, and 
it is more conspicuous than the dwarfs of the same colour. 3) The CaII 
H and K lines near 3950$\AA$, which are sensitive to $[M/H]$.  

The $(g-r)_{0}$ versus $(u-g)_{0}$ colour-colour diagram of \citet{Helmi03}
for 19,000 stars brighter than $r_{0}=19$ and whose photometric errors in all bands 
are less than 0.05 show a clear offset for nine $SPS$ giants. The authors used 
the well-defined stellar locus to derive a principal axes coordinate system 
($P_{1}$, $P_{2}$), where $P_{1}$ lies parallel to the stellar locus and $P_{2}$ measures 
the distance from it. The origin is chosen to coincide with the highest stellar 
density ($(u-g)_{0}=1.21$, $(g-r)_{0}=0.42$). Since the objects of interest 
occur in a relatively narrow colour range, they restricted their work to 
$1.1\leq(u-g)_{0}\leq 2$ and $0.3\leq(g-r)_{0}\leq0.8$. This procedure 
yields 

\begin{eqnarray}
P_{1}=0.910(u-g)_{0}+0.415(g-r)_{0}-1.28,\nonumber\\
P_{2}=-0.415(u-g)_{0}+0.910(g-r)_{0}+0.12,
\end{eqnarray}
The position of the locus in the $r$ versus $P_{2}$ colour-magnitude diagram 
depends on the $r$ magnitudes, i.e. the median $P_{2}$ colour becomes redder 
at the faint end. \citet{Helmi03} corrected for this effect using a linear 
$P_{2}$ versus $r$ fit (the corrections varies from -0.03 to 0.05 mag). 
Thus, they defined the colour $s$ that is normalized such that its error is 
approximately equal to the mean photometric error in a single band. 
They obtained

\begin{equation}
s=-0.249u_{0}+ 0.794g_{0}–-0.555r_{0}+0.240.
\end{equation}
The diagram $r$ versus $s$ show a symmetrical distribution for the star 
sample in question and a clear offset for the giants.

Based on the $s$ colour distribution of the $SPS$ giants and the overall $s$ 
colour distribution, \citet{Helmi03} selected metal-poor giants as stars with 
$r_{0}<19$, $-0.1<P_{1}<0.6$ for $1.1\leq(u-g)_{0}\leq2$ and 
$0.3\leq(g-r)_{0}\leq0.8$, and $|s|>m_{s}+0.05$, where $m_{s}$ is the 
median value of $s$ in appropriately chosen subsamples.  

Thus, the 60 total giants were excluded from the sample.

\begin{figure}
\begin{center}
\includegraphics[angle=0, width=80mm, height=78.5mm]{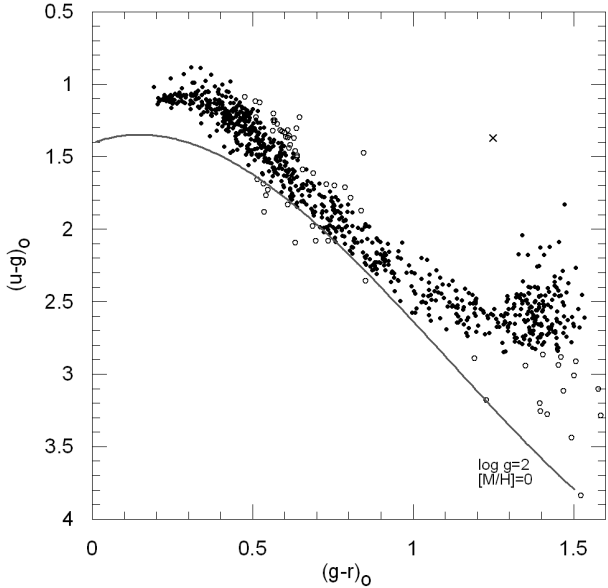}
\caption[] {The position of giants ($\circ$) and dwarfs ($\bullet$)
in the $ (u-g)_{0}-(g-r)_{0}$ two colour diagram, identified by the
procedures of \citet{Ivezic07} and \citet{Helmi03}. The cross 
($\times$) corresponds to the unidentified star. The solid line is 
based on the \citet{Kurucz79} stellar model for $\log g=2$ and 
$[M/H]=0$ dex.}
\end{center}
\end{figure}

\subsection{Final sample}
We omitted one final standard from the sample due to its position in
the two-colour diagrams (Fig. 4). The final sample includes
825 stars. Their {\em 2MASS} $JHK_{s}$, {\em SDSS} $gri$ and {\em
BVRI} data are given Table 1. Also, the $(u-g)_{0}-(g-r)_{0}$ and 
$(g-r)_{0}-(r-i)_{0}$ two--colour diagrams are plotted in Fig. 5. 
The errors for the magnitudes in $B$, $V$, $R$, $I$, $g$, $r$, 
$i$, $J$, $H$, and $K_{s}$ are presented in Table 2 and Fig. 6. The 
colour intervals covered by the standards are $0.30<(B-V)_{0}<1.70$, 
$0.23<(R-I)_{0}<1.78$, $0.18<(g-r)_{0}<1.54$, $0<(r-i)_{0}<1.65$, 
$0<(J-H)_{0}<0.98$, and $-0.23<(H-K_{s})_{0}<0.63$.

\begin{figure}
\begin{center}
\includegraphics[angle=0, width=80mm, height=146.9mm]{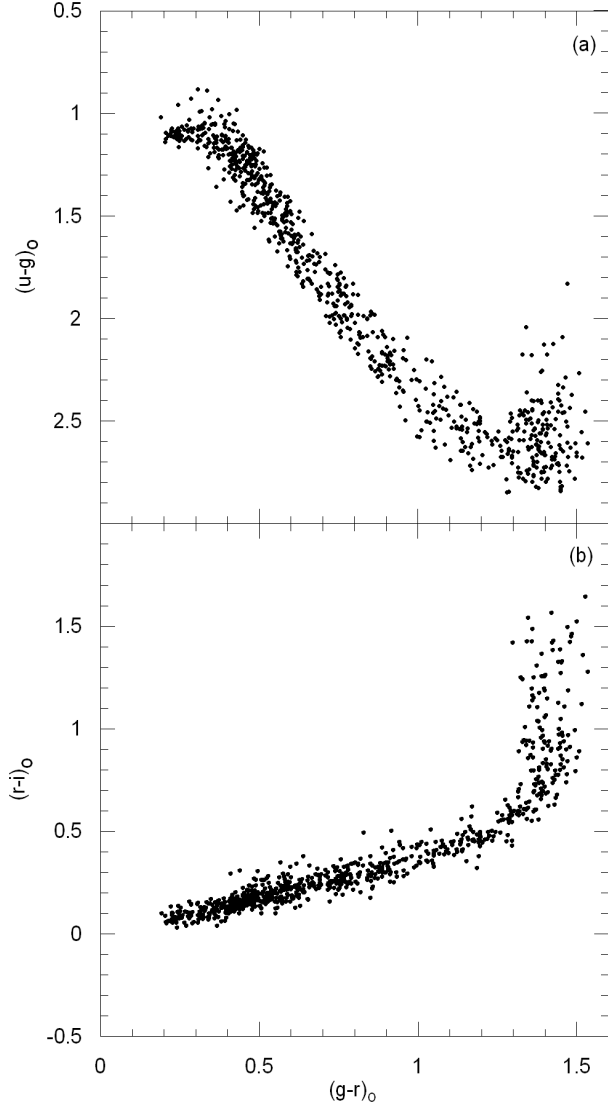}
\caption[] {Two-colour diagrams of the sample (825 dwarfs). (a) the $(u-g)_{
0}-(g-r)_{0}$ and (b) the $(g-r)_{0}-(r-i)_{0}$ diagram.}
\end{center}
\end{figure}

\begin{figure}
\begin{center}
\includegraphics[angle=0, width=80mm, height=196.4mm]{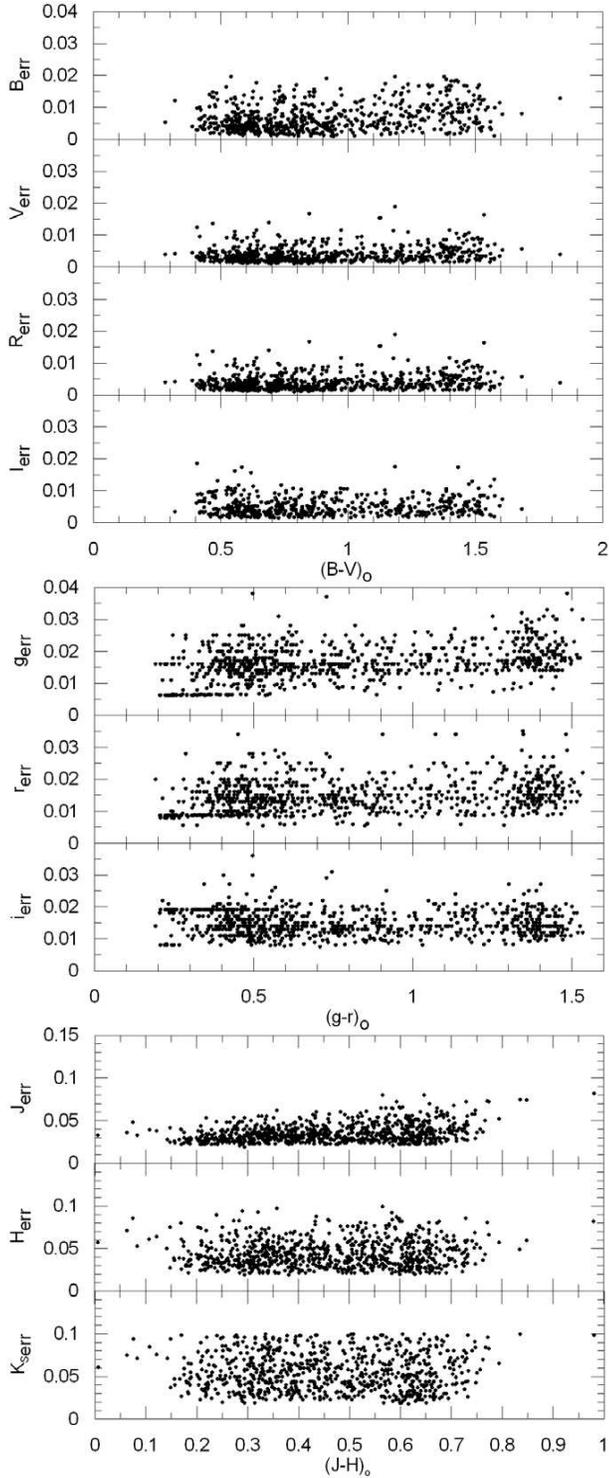}
\caption[] {The error distributions for Johnson-Cousins $BVRI$, {\em SDSS} $
gri$, and {\em 2MASS} $JHK_{s}$.}
\end{center}
\end{figure}

\begin{table}
\center \caption{Mean errors and standard deviations for the filters
of Johnson-Cousins, {\em SDSS}, and {\em 2MASS} photometries.}
\begin{tabular}{cccc}
\hline
 Filter & Mean error & $s$ & Photometry \\
\hline
         B & 0.006 & $\pm$0.005 & {\em BVRI} \\
         V & 0.004 &      0.003 &            \\
         R & 0.005 &      0.005 &            \\
         I & 0.004 &      0.004 &            \\
         g & 0.016 &      0.005 & {\em SDSS} \\
         r & 0.015 &      0.005 &            \\
         i & 0.015 &      0.004 &            \\
         J & 0.036 &      0.011 & {\em 2MASS}\\
         H & 0.044 &      0.016 &            \\
   $K_{s}$ & 0.058 &      0.024 &            \\

\hline
\end{tabular}
\end{table}

The $g_{0}$ histogram presented in Fig. 7 shows that our sample
includes stars of different apparent magnitudes in a large range:
$14<g_{0}<20$. On the other hand, the $(g-r)_{0}$ colour
histogram in the same figure indicates a multi-modal distribution.
This is very important because if the star sample consists of a
combination of different population types and metallicities, then
different transformation equations should be derived. We evaluated
the metallicity of stars by the procedure given in
\citet{Karaali03}. This procedure, defined for stars with $0.10 <
(g-r)_{0}\leq0.95$, provides metallicities in the interval $-2.7
\leq[M/H]\leq +0.1$ dex and is based on the calibration of 
the metallicity determined spectroscopically, where $\delta_{0.43}$ 
is the ultraviolet excess of a star relative to a Hyades star of the 
same $(g-r)_{0}$, reduced to the colour $(g-r)_{0}=0.43$ which 
corresponds to $(B-V)_{0}=0.60$ in the {\em UBV}–-system:
$[M/H]=0.10-3.54\delta_{0.43}-39.63\delta_{0.43}^{2}+63.51\delta_{0.43}^{3}.$
The restriction of $(g-r)_{0}$ is due to the colour 
range of the sample used for the calibration. Red stars, 
$(g-r)_{0}>0.95$, are old thin disc dwarfs with a mean metallicity 
$[M/H]=-0.1\pm0.3$ dex \citep{Cox00}. Hence we adopted the metallicity 
range for these stars as $[M/H]>-0.4$ dex. We assumed stars with $(g-r)_{0}>0.95$
to be metal-rich (Table 3). Fig. 8 shows that our sample covers stars with
metallicities down to $[M/H]=-3$ dex. The number of metal-poor
stars ($[M/H] \leq -1.2$ dex) is not negligible. Therefore, we
separated the sample of stars into three metallicity categories:
metal-rich stars ($[M/H]>-0.4$ dex), intermediate-metallicity stars
($-1.2 < [M/H] \leq -0.4$ dex), and metal-poor stars ($-3<[M/H]
\leq -1.2$ dex). Transformation equations for each set are
evaluated.

\begin{figure}
\begin{center}
\includegraphics[angle=0, width=80mm, height=109.0mm]{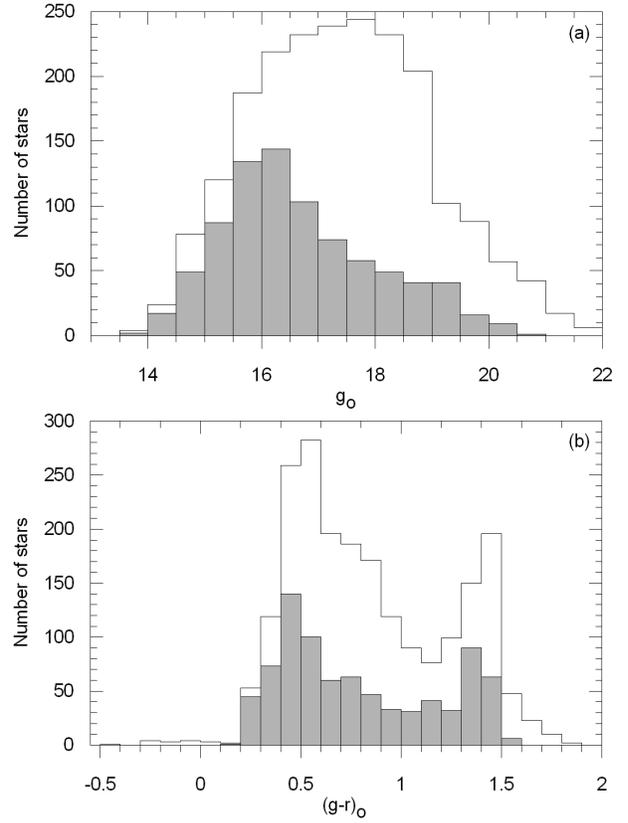}
\caption[] {(a) $g_{0}$ apparent magnitude histogram, and (b)
$(g-r)_{0}$ colour histogram for the ``AAA'' star sample (black
area) and for the stars found in all three photometries (white
area).}
\end{center}
\end{figure}

\begin{figure}
\begin{center}
\includegraphics[angle=0, width=80mm, height=54.6mm]{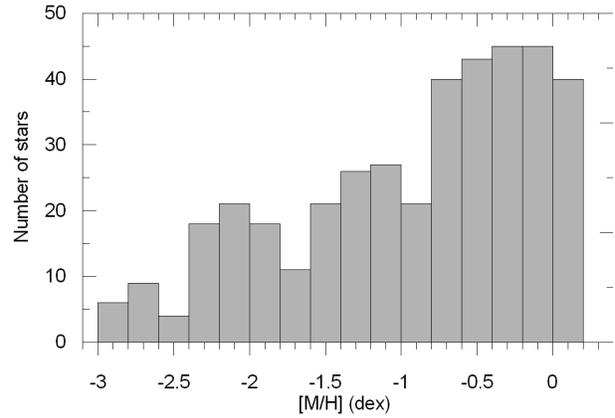}
\caption[] {Metallicity distribution of the sample.}
\end{center}
\end{figure}

\begin{table}
\center \caption{Metallicity distribution of the sample. Stars with
$(g-r)_{0}>0.95$ mag were assumed to have a metallicity of
$-0.4<[M/H]$ dex.}
\begin{tabular}{cc}
\hline
Metallicity (dex) & Number of stars \\
\hline
$-0.4<[M/H]$          & 505 \\
$-1.2<[M/H]\leq -0.4$ & 131 \\
$-3.0<[M/H]\leq -1.2$ & 189 \\
\hline
\end{tabular}
\end{table}

\section{Results}

\subsection{Transformations between {\em 2MASS} and Johnson-Cousins photometry}

We used the following general equations and derived four sets of
transformations between {\em 2MASS} and Johnson--Cousins {\em BVRI}. The
sets of transformations consist of: 1) transformations for the whole
sample (825 stars), 2) transformations for metal-rich stars, 3)
transformations for intermediate metallicity stars, and 4)
transformations for metal-–poor stars. The definition of the last
three sub-samples are already given in the previous section. The
general equations are:

\begin{equation}
(V-–J)_{0}=a_{1}(B-V)_{0}+b_{1}(R-I)_{0}+c_{1},
\end{equation}

\begin{equation}
(V–-H)_{0}=a_{2}(B-V)_{0}+b_{2}(R-I)_{0}+c_{2},
\end{equation}

\begin{equation}
(V–-K_{s})_{0}=a_{3}(B-V)_{0}+b_{3}(R-I)_{0}+c_{3}.
\end{equation}

The numerical values of the coefficients $a_{i}$, $b_{i}$ and
$c_{i}$ ($i$=1, 2, 3) for the four sets are given in Table 4. The
fourth and fifth numbers in each column are the squared correlation
coefficient and the standard deviation for the colour indicated at
the top of the column. There are differences between the values of
the coefficients evaluated for the largest sample (825 stars) and
the three sub samples. However, the coefficients for metal-rich
stars are close to those corresponding to the total sample. The same
similarity can be seen for the intermediate metallicity stars and
the metal–-poor stars. The metallicity distribution of these sub
samples ($-0.4<[M/H]$, $-1.2<[M/H]\leq-0.4$, and 
$-3<[M/H]\leq-1.2$ dex) remind us of the metallicity ranges
depending on stellar location in thin or thick discs or the halo,
respectively. That is, the transformations are luminosity and
metallicity dependent. On the other hand, the coefficients $a_{i}$
and $b _{i}$ for the same equation are numerically comparable,
suggesting that the transformations are two-colour dependent. Thus,
if we derived transformations depending on a single colour, accuracy
would be lost. In some works \citep[cf.,][for example]{Jordi06}
transformations were dependent on a single colour and the authors
compensate by using a step function.

\begin{table*}
\center \caption{Coefficients $a_{i}$, $b_{i}$, and $c_{i}$ for the
transformation equations (10), (11), and (12), in column matrix form
for the four star categories. The subscript $i$=1, 2, and 3
correspond to the same number that denotes the columns. Numerical
values in the fourth and fifth lines of each category are the
squared correlation coefficients ($R^{2}$) and the standard deviations 
($s$), respectively.}
\begin{tabular}{ccccc}
\hline
           &       &        (1) &        (2) &        (3) \\
\hline
    Category &            & $(V-J)_{o}$ & $(V-H)_{o}$ & $(V-K_{s})_{o}$ \\
\hline
Total sample &   $a_{i}$&  1.210 $\pm$ 0.032 &  1.816 $\pm$ 0.039 &  1.896 $\pm$ 0.044 \\
           &     $b_{i}$&  1.295 $\pm$ 0.038 &  1.035 $\pm$ 0.046 &  1.131 $\pm$ 0.052 \\
           &     $c_{i}$& -0.046 $\pm$ 0.014 &  0.016 $\pm$ 0.017 & -0.004 $\pm$ 0.019 \\
           &    $R^{2}$ &      0.983 &      0.982 &      0.980 \\
           &        $s$ &      0.098 &      0.119 &      0.133 \\
\hline
$-0.4<[M/H]$ &   $a_{i}$&  1.180 $\pm$ 0.042 &  1.815 $\pm$ 0.050 &  1.878 $\pm$ 0.058 \\
           &     $b_{i}$&  1.346 $\pm$ 0.045 &  1.062 $\pm$ 0.052 &  1.165 $\pm$ 0.061 \\
           &     $c_{i}$& -0.051 $\pm$ 0.023 & -0.015 $\pm$ 0.027 & -0.018 $\pm$ 0.032 \\
           &    $R^{2}$ &      0.985 &      0.984 &      0.981 \\
           &        $s$ &      0.103 &      0.121 &      0.142 \\
\hline
$-1.2<[M/H]\leq-0.4$ & $a_{i}$ &  1.557 $\pm$ 0.111 &  2.109 $\pm$ 0.157 &  2.031 $\pm$ 0.150 \\
           &     $b_{i}$       &  0.461 $\pm$ 0.205 &  0.612 $\pm$ 0.290 &  0.878 $\pm$ 0.277 \\
           &     $c_{i}$       &  0.049 $\pm$ 0.049 & -0.016 $\pm$ 0.069 &  0.004 $\pm$ 0.066 \\
           &    $R^{2}$ &      0.902 &      0.894 &      0.906 \\
           &        $s$ &      0.080 &      0.113 &      0.108 \\
\hline
$-3.0<[M/H]\leq -1.2$ & $a_{i}$ & 1.542 $\pm$ 0.081 &  1.920 $\pm$ 0.108 &  2.044 $\pm$ 0.123 \\
           &    $b_{i}$        &  0.447 $\pm$ 0.156 &  0.845 $\pm$ 0.207 &  0.974 $\pm$ 0.237 \\
           &    $c_{i}$        &  0.095 $\pm$ 0.036 &  0.055 $\pm$ 0.048 & -0.022 $\pm$ 0.054 \\
           &    $R^{2}$ &      0.943 &      0.942 &      0.936 \\
           &        $s$ &      0.084 &      0.112 &      0.128 \\
\hline
\end{tabular}
\end{table*}

\subsection{Transformations between {\em 2MASS} and {\em SDSS}}

The transformations between {\em 2MASS} and {\em SDSS} have similar general
equations, given below:

\begin{equation}
(g-–J)_{0}=d_{1}(g-r)_{0}+e_{1}(r-i)_{0}+f_{1},
\end{equation}

\begin{equation}
(g–-H)_{0}=d_{2}(g-r)_{0}+e_{2}(r-i)_{0}+f_{2},
\end{equation}

\begin{equation}
(g-–K_{s})_{0}=d_{3}(g-r)_{0}+e_{3}(r-i)_{0}+f_{3}.
\end{equation}
The numerical values of the coefficients $d_{i}$, $e_{i}$ and
$f_{i}$ ($i= 1, 2, 3$) for the four sets defined above, are given in
Table 5. The transformations between {\em 2MASS} and {\em SDSS} are
also luminosity, metallicity, and two-colour dependent for the same
reasons explained in Section 3.1.

\begin{table*}
\center
\caption{Coefficients $d_{i}$, $e_{i}$, and $f_{i}$ for the
transformation equations (13), (14), and (15), in column matrix form
for the four star categories. The subscript $i$=1, 2, and 3
correspond to the same number that denotes the columns. Numerical
values in the fourth and fifth lines of each category are the
squared correlation coefficients ($R^{2}$) and the standard deviations ($s$),
respectively.}
\begin{tabular}{ccccc}
\hline
           &            &        (1) &        (2) &        (3) \\
\hline
   Category &           & $(g-J)_{o}$ & $(g-H)_{o}$ & $(g-K_{s})_{o}$ \\
\hline
Total sample& $d_{i}$          &  1.379 $\pm$ 0.015 &  1.849 $\pm$ 0.021 &  1.907 $\pm$ 0.023 \\
           &  $e_{i}$          &  1.702 $\pm$ 0.019 &  1.536 $\pm$ 0.025 &  1.654 $\pm$ 0.028 \\
           &  $f_{i}$          &  0.518 $\pm$ 0.007 &  0.666 $\pm$ 0.010 &  0.684 $\pm$ 0.011 \\
           &    $R^{2}$ &      0.994 &      0.991 &      0.990 \\
           &        $s$ &      0.083 &      0.115 &      0.126 \\
\hline
$-0.4<[M/H]$ & $d_{i}$         &  1.361 $\pm$ 0.016 &  1.823 $\pm$ 0.022 &  1.881 $\pm$ 0.024 \\
           &   $e_{i}$         &  1.724 $\pm$ 0.019 &  1.561 $\pm$ 0.026 &  1.675 $\pm$ 0.028 \\
           &   $f_{i}$         &  0.521 $\pm$ 0.009 &  0.670 $\pm$ 0.013 &  0.692 $\pm$ 0.014 \\
           &    $R^{2}$ &      0.995 &      0.993 &      0.992 \\
           &        $s$ &      0.080 &      0.111 &      0.121 \\
\hline
$-1.2<[M/H]\leq-0.4$ & $d_{i}$         &  1.536 $\pm$ 0.102 &  1.792 $\pm$ 0.134 &  1.790 $\pm$ 0.143 \\
           &    $e_{i}$                &  1.400 $\pm$ 0.215 &  2.092 $\pm$ 0.281 &  2.272 $\pm$ 0.301 \\
           &    $f_{i}$                &  0.488 $\pm$ 0.028 &  0.584 $\pm$ 0.037 &  0.628 $\pm$ 0.039 \\
           &    $R^{2}$ &      0.928   &      0.924 &      0.918 \\
           &        $s$ &      0.085   &      0.112 &      0.120 \\
\hline
$-3.0<[M/H]\leq-1.2$ &  $d_{i}$&  1.435 $\pm$ 0.061 &  1.711 $\pm$ 0.075 &  1.741 $\pm$ 0.086 \\
           &    $e_{i}$        &  1.769 $\pm$ 0.137 &  2.339  $\pm$ 0.169 & 2.640 $\pm$ 0.192 \\
           &    $f_{i}$        &  0.481 $\pm$ 0.022 &  0.598 $\pm$ 0.027 &  0.583 $\pm$ 0.031 \\
           &    $R^{2}$ &      0.960 &      0.960 &      0.954 \\
           &        $s$ &      0.088 &      0.108 &      0.123 \\
\hline
\end{tabular}
\end{table*}

\subsection{Residuals}

We compared the observed colours and those evaluated via Eqs. (10)
–-- (15). The mean of the residuals are smaller than a thousandth.
These residuals can be found in Table 6. The standard
deviations, also given in Table 6, are close to 0.1 for all colours.
The residuals are plotted versus observed $(B-V)_{0}$ or $(g
–-r)_{0}$ colours in Fig. 9. Although the number of stars are not
equal in each bin, there is no systematic deviation from the zero
point in any panel. However, the ranges of the residuals for
different colours are not the same. Those resulting from Eqs. (10)
and (13) are the smallest. The ranges of the residuals are larger
for the longer-wavelength magnitudes. That is, for the $J$
magnitude, the residual $\Delta(V-J)$ and $\Delta(g-J)$ lies between
-0.2 and +0.2, with a few exceptions, whereas $\Delta(V-H)$,
$\Delta(g-H)$, $\Delta(V-K_{s})$ and $\Delta(g-K_{s})$ extend down
to -0.4 and up to +0.4. This indicates that the $J$ (absolute)
magnitudes evaluated via the transformations given above would be
more accurate.

We did not show the plots of actual transformations. However, 
the comparison of the residuals, for the entire sample, for 
two--colours and one--colour transformations in Fig. 10 show that 
our transformations are much better than those for the one--colour ones. 
The scatter for one--colour transformations is much larger than that 
for two--colours and additionally there is a systematic deviation the 
one-colour transformations.

\begin{figure*}
\begin{center}
\includegraphics[angle=0, width=160mm, height=130.9mm]{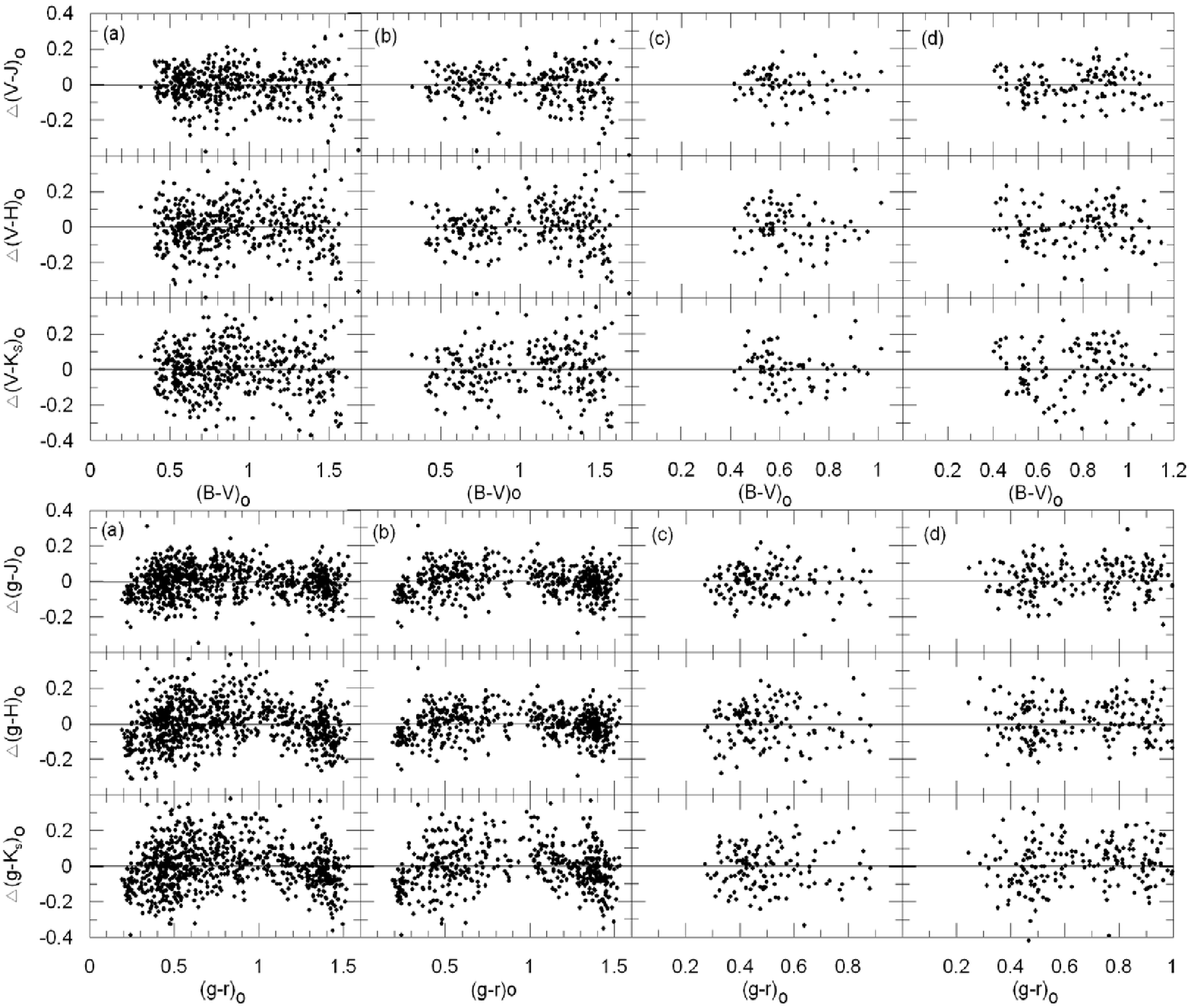}
\caption[] {Colour residuals, for four star categories. The notation
used is $\Delta$(colour) = (evaluated colour) -- (measured colour).
The four categories are (from left to right) (a) the entire sample,
(b) high metallicity (metal-rich), (c) intermediate metallicity, and 
(d) low metallicity (metal-poor).}
\end{center}
\end{figure*}

\begin{table*}
\center
\caption{Averages and standard deviations ($s$) for differences
between the measured and calculated colours (residuals) for six
colours in all four star categories. The notation used is
$\Delta$(colour) = (evaluated colour) –- (measured colour).}
\begin{tabular}{cccccccc}
\hline
Category  &            & $\Delta (V-J)_o$ & $\Delta (V-H)_o$ & $\Delta (V-K_{s
})_o$ & $\Delta (g-J)_o$ & $\Delta (g-H)_o$ & $\Delta (g-K_{s})_o$ \\
\hline
Total sample &    average &    -0.0002 &    -0.0004 &     0.0001 &     0.0008 &     0.0002 &     0.0006 \\
           &          $s$ &      0.098 &      0.119 &      0.133 &      0.083 &      0.114 &      0.126 \\
$-0.4<[M/H]$ &    average &    -0.0002 &     0.0000 &    -0.0001 &    -0.0002 &    -0.0004 &    -0.0007 \\
           &          $s$ &      0.103 &      0.121 &      0.141 &      0.080 &      0.111 &      0.121 \\
$-1.2<[M/H]\leq-0.4$ &    average &     0.0004 &     0.0001 &    -0.0004 &   0.0000 &     0.0003 &     0.0001 \\
           &          $s$ &      0.079 &      0.112 &      0.107 &      0.085&      0.111 &      0.119 \\
$-3.0<[M/H]\leq-1.2$ &    average &     0.0001 &     0.0001 &     0.0000 &-0.0003 &    -0.0004 &    0.0002 \\
           &          $s$ &      0.084 &      0.111 &      0.127 &      0.087 &      0.107 &      0.122 \\
\hline
\end{tabular}
\end{table*}

\begin{table*}
\center
{\tiny
\caption{Coefficients for the inverse transformation equations for four star 
categories. $\alpha_{i}$, $\beta_{i}$, and $\gamma_{i}$ ($i$ = 1, 2, 3 and 4) 
correspond to Eqs. (16), (17), (18) and (19), respectively. The numerical 
values of the coefficients are indicated on the same line of the corresponding 
star category.}
\begin{tabular}{lcccccccccc}
\hline
Johnson–-Cousins system& \multicolumn{5}{c}{$(B-V)_{0}=\alpha_{1}(J-H)_{0}+\beta_{1}(H-K_{s})_{0}+\gamma_{1}$} &\multicolumn{5}{c}{$(R-I)_{0}=\alpha_{2}(J-H)_{0}+\beta_{2}(H-K_{s})_{0}+\gamma_{2}$} \\
\hline
  Category &  $\alpha_{1}$ &  $\beta_{1}$ &    $\gamma_{1}$ & $R^{2}$ & $s$ & $\alpha_{2}$ &  $\beta_{2}$ &    $\gamma_{2}$ & $R^{2}$ & $s$\\
\hline
Total Sample & 1.622$\pm$0.032& 0.912$\pm$0.051 & 0.044$\pm$0.015 & 0.845 & 0.120 & 0.954$\pm$0.028 & 0.593$\pm$0.050 & 0.025$\pm$0.013 & 0.755 & 0.101 \\
$-0.4<[M/H]$ &1.640$\pm$0.044 & 1.033$\pm$0.075 & 0.050$\pm$0.022 & 0.855 & 0.125 & 1.027$\pm$0.040 & 0.658$\pm$0.080 &-0.003$\pm$0.020 & 0.772 & 0.117 \\
$-1.2<[M/H]\leq-0.4$ &1.103$\pm$0.074 & 0.486$\pm$0.091 & 0.228$\pm$0.029 & 0.665 & 0.077 & 0.521$\pm$0.054 & 0.311$\pm$0.066 &0.179$\pm$0.021 & 0.546 & 0.050 \\
$-3.0<[M/H]\leq-1.2$ & 1.276$\pm$0.056 & 0.541$\pm$0.066 & 0.173$\pm$0.025 & 0.782 & 0.088 & 0.608$\pm$0.038 & 0.322$\pm$0.051 &0.172$\pm$0.017 & 0.712 & 0.054 \\
\hline
{\em SDSS} system& \multicolumn{5}{c}{$(g-r)_{0}=\alpha_{3}(J-H)_{0}+\beta_{3}(H-K_{s})_{0}+\gamma_{3}$} &\multicolumn{5}{c}{$(r-i)_{0}=\alpha_{4}(J-H)_{0}+\beta_{4}(H-K_{s})_{0}+\gamma_{4}$} \\
\hline
 Category &  $\alpha_{3}$ &  $\beta_{3}$ &    $\gamma_{3}$ & $R^{2}$ & $s$ & $\alpha_{4}$ &  $\beta_{4}$ &    $\gamma_{4}$ & $R^{2}$ & $s$\\
\hline
Total Sample & 1.951$\pm$0.032& 1.199$\pm$0.050 &-0.230$\pm$0.015 & 0.879 & 0.135 & 0.991$\pm$0.026 & 0.792$\pm$0.042 &-0.210$\pm$0.012 & 0.760 & 0.107 \\
$-0.4<[M/H]$ & 1.991$\pm$0.040& 1.348$\pm$0.066 &-0.247$\pm$0.019 & 0.900 & 0.136 & 1.000$\pm$0.036 & 1.004$\pm$0.064 &-0.220$\pm$0.017 & 0.779 & 0.120 \\
$-1.2<[M/H]\leq-0.4$ & 1.217$\pm$0.078& 0.491$\pm$0.091 &0.050$\pm$0.030 & 0.663 & 0.083 & 0.600$\pm$0.035 & 0.268$\pm$0.040 &-0.049$\pm$0.013 & 0.708 & 0.037 \\
$-3.0<[M/H]\leq-1.2$ & 1.422$\pm$0.065& 0.600$\pm$0.076 &-0.003$\pm$0.029 & 0.749 & 0.099 & 0.609$\pm$0.030 & 0.279$\pm$0.035 &-0.047$\pm$0.013 & 0.728 & 0.045 \\
\hline
\end{tabular}  
}
\end{table*}

\subsection{Inverse transformation formulae}
One may need to transform {\em 2MASS} data to Johnson–-Cousins system or 
{\em SDSS} system. Hence, we derived the inverse transformation formulae 
of Eqs. (10)$-–$(12) and (13)$-$(15) as follows. We reduced the Eqs. (10)$-$(12) 
to two equations by eliminating the $V$ magnitude. The solution of these 
equations gives the Johnson-Cousins colours $(B-V)_{0}$ and $(R-I)_{0}$ as 
a function of $(J-H)_{0}$ and $(H-K_{s})_{0}$. The general equations are:
\begin{equation}
(B-V)_{0}=\alpha_{1}(J-H)_{0}+\beta_{1}(H-K_{s})_{0}+\gamma_{1},
\end{equation}
\begin{equation}
(R-I)_{0}=\alpha_{2}(J-H)_{0}+\beta_{2}(H-K_{s})_{0}+\gamma_{2}.
\end{equation}
We derived the following general equations for the {\em SDSS} colours, 
i.e. $(g-r)_{0}$ and $(r-i)_{0}$, by applying the same procedure to the 
Eqs. (13)$-$(15).
\begin{equation}
(g-r)_{0}=\alpha_{3}(J-H)_{0}+\beta_{3}(H-K_{s})_{0}+\gamma_{3},
\end{equation}
\begin{equation}
(r-i)_{0}=\alpha_{4}(J-H)_{0}+\beta_{4}(H-K_{s})_{0}+\gamma_{4}.
\end{equation}
The numerical values of the coefficients $\alpha_{i}$, $\beta_{i}$, and 
$\gamma_{i}$ ($i$=1, 2, 3, 4) for the four sets mentioned above are given 
in Table 7 .

\section{Conclusion}

We have presented the colour transformations for the conversion of
the {\em 2MASS} photometric system into the Johnson–-Cousins $BVRI$
system and further into the {\em SDSS} $gri$ system. We have added 
some constraints to the dataset in addition to those of other authors 
who derived transformations between {\em SDSS} $ugriz$ and other systems 
in order to obtain the most accurate transformations possible. The overall 
constraints used were as follows: 1) the data were de-reddened, 2) giants 
have been identified and excluded from the sample, 3) sample stars have 
been selected by the quality of the data, 4) transformations have been 
derived for sub-samples of different metallicity and populations type, 
and 5) transformations are two colour dependent. The constraints described 
in items (1), (2), and (3) are new. Constrain (3) is especially important 
for {\em 2MASS} data because the inherent errors are larger relative to 
those in others photometries       

The squared correlation coefficients ($R^{2}$) for the
transformations carried out for the four categories (i.e., the
entire sample, the metal-rich stars, the intermediate metallicity
stars, and the metal-poor stars) are rather high (see Tables 4 and
5). The smallest of those is $R^{2}=0.894$ for the colour
$(V-H)_{0}$. All others lie between 0.900 and 0.995. The standard
deviation is 0.1. The coefficients of the colour terms in the
same transformation equation are compatible (see Table 4 and Table
5) which indicate that the transformations are two-colour dependent. 
On the other hand, there are differences between the corresponding
transformation coefficients for the four categories. This is most
conspicuous for metal-rich and metal-poor stars. That is, our
transformations are metallicity dependent. Additionally, the
metallicity ranges of the three sub samples, $-0.4<[M/H]$,
$-1.2<[M/H]\leq- 0.4$, and $-3<[M/H]\leq-1.2$ dex, correspond to
that for a thin disc, thick disc, and halo (i.e., the
transformations are also luminosity dependent).

The mean of the residuals (i.e. the colour differences between those
measured and those calculated) are less than $\pm 0.001$ mag (see Table 6).
The deviations of the measured colours from the calculated ones are 
small for $(V-J)_{0}$, $\pm0.2$, but relatively larger for redder 
colours ($\pm 0.3$ for $(V-H)_{0}$ and $(V–-K_{s})_{0}$) for most of 
the stars (Fig. 9). Thus, the $J$ absolute magnitudes will yield the 
best estimations. \citet{Davenport06} did not give any residuals for 
{\em 2MASS} data and thus we cannot compare their transformations with 
ours. However, the deviations of the measured colours from those calculated 
are not larger than those claimed by the authors cited above \citep[cf. see
Figures 4 and 6 of][]{Jordi06}. If we take into account that the
errors for {\em 2MASS} data are larger than the errors of the
photometries used in other recent transformations (such as between
Johnson-Cousins $UBVRI$ and {\em SDSS} $ugriz$ data), and that the
data are non-simultaneous, we can say that our transformations are
quite accurate.

\begin{figure}
\begin{center}
\includegraphics[angle=0, width=80mm, height=53.1mm]{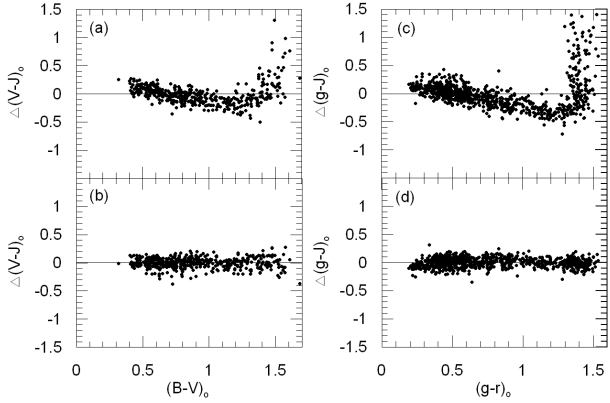}
\caption[] {$\Delta(V-J)_{0}$ and $\Delta(g-J)_{0}$ colour residuals 
versus observed $(B-V)_{0}$ or $(g-r)_{0}$, as an example. (a) and 
(c) for one--colour transformations, (b) and (d) for two--colour 
transformations.}
\end{center}
\end{figure}

\section{Acknowledgments}

We would like to thank James Davenport, the referee, for his useful 
and constructive comments concerning the manuscript. This research 
used the facilities of the Canadian Astronomy Data Centre operated 
by the National Research Council of Canada with the support of the 
Canadian Space Agency.

Funding for the {\em SDSS} and {\em SDSS-II} has been provided by
the Alfred P. Sloan Foundation, the Participating Institutions, the
National Science Foundation, the U.S. Department of Energy, the
National Aeronautics and Space Administration, the Japanese
Monbukagakusho, the Max Planck Society, and the Higher Education
Funding Council for England. The {\em SDSS} Web Site is
http://www.sdss.org/. {\em SDSS} is managed by the Astrophysical
Research Consortium for the participating institutions. 

This publication makes use of data products from the Two Micron All
Sky Survey, which is a joint project of the University of
Massachusetts and the Infrared Processing and Analysis
Center/California Institute of Technology, funded by the National
Aeronautics and Space Administration and the National Science
Foundation.

This research has made use of the SIMBAD, NASA's Astrophysics Data
System Bibliographic Services and the NASA/IPAC Extragalactic
Database (NED) which is operated by the Jet Propulsion Laboratory,
California Institute of Technology, under contract with the National
Aeronautics and Space Administration.

\end {document}